\documentclass{cimento}

\usepackage{graphicx}

\title{Viability of SUSY-GUT paradigm after LHC}
\author{Marco~Chianese}
\instlist{\inst{}Dipartimento di Fisica, Universit\`a di Napoli ``Federico II", Complesso Univ. Monte S. Angelo, Via Cinthia, I-80126 Napoli, Italy.  \inst{}  INFN, Sezione di Napoli, Complesso Univ. Monte S. Angelo, Via Cinthia, I-80126 Napoli, Italy.}

\begin{document}

\maketitle

\begin{abstract}
The SUSY-GUT paradigm is the most promising scenario for the physics beyond the Standard Model. After the LHC run I, it is of interest  to reanalyze the room still remaining for SUSY-GUT inspired models and to study the limits on the SUSY mass spectrum. Assuming one step unification of gauge couplings, under some {\it natural} requirements we have obtained the energy upper bound for the observation of SUSY phenomenology. We found that in the SUSY-GUT framework the mass of lightest gluino or Higgsino cannot be larger than about 20~TeV.
\end{abstract}

\section{Introduction}

The recent detection of the Higgs particle at LHC certainly represents a milestone which has once again confirmed the predictive effectiveness of the Standard Model (SM) of strong and electroweak interactions. On the other hand, the missing evidence of supersymmetric particles at LHC has greatly constrained Supersymmetry (SUSY) and Supersymmetric Grand Unified Theories (SUSY-GUT). In our analysis \cite{Berezhiani:2015vea} we reanalyze the level of viability of SUSY-GUT inspired models, showing that under some {\it naturalness} principles the {\it energy upper bound} for SUSY phenomenology is about 20~TeV. In particular, we require:
\begin{itemize}
\item One step unification of the three gauge couplings $SU(3) \times SU(2) \times U(1)$ at a single energy scale, denoted as $M_{\rm GUT}$, without the presence of intermediate symmetry breaking scales ({\it $SU(5)$ bottleneck}). Below $M_{\rm GUT}$ the theory spontaneously breaks to the Minimal Supersymmetric Standard Model (MSSM).
\item Agreement with masses of third generation fermions and with the experimental bound on proton decay $p \rightarrow e^+\pi^0$ \cite{Nishino:2012ipa}, which provides the constraint $M_{\rm GUT}/\sqrt{\alpha_{\rm GUT}} \geq 3 \cdot 10^{16}$~GeV \cite{Hisano:2012wq}.
\item No {\it ad hoc} small parameters and no {\it ad hoc} fine tuning among parameters in the SUSY-GUT model that holds above $M_{\rm GUT}$. This {\it naturalness} requirement implies that all the couplings at $M_{\rm GUT}$ have to be ${\cal{O}}(1)$.
\end{itemize}

\section{Method and result}

\begin{figure}[t!]
\centering
$\chi_\Sigma=1$ \hskip41.mm $\chi_\Sigma=10$\\
\vspace{1.mm}
\includegraphics[scale=0.22]{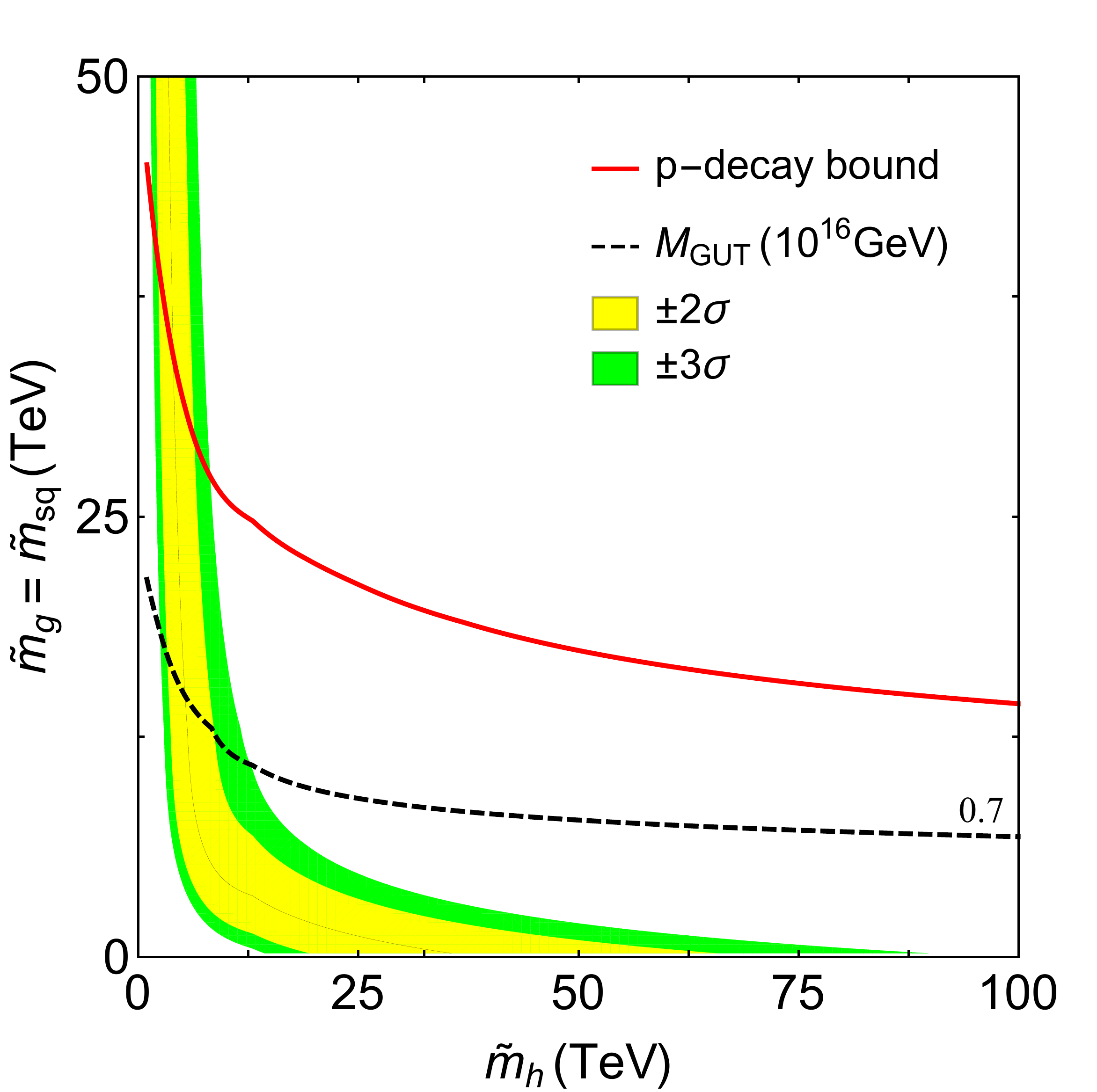}
\hspace{7.mm}
\includegraphics[scale=0.22]{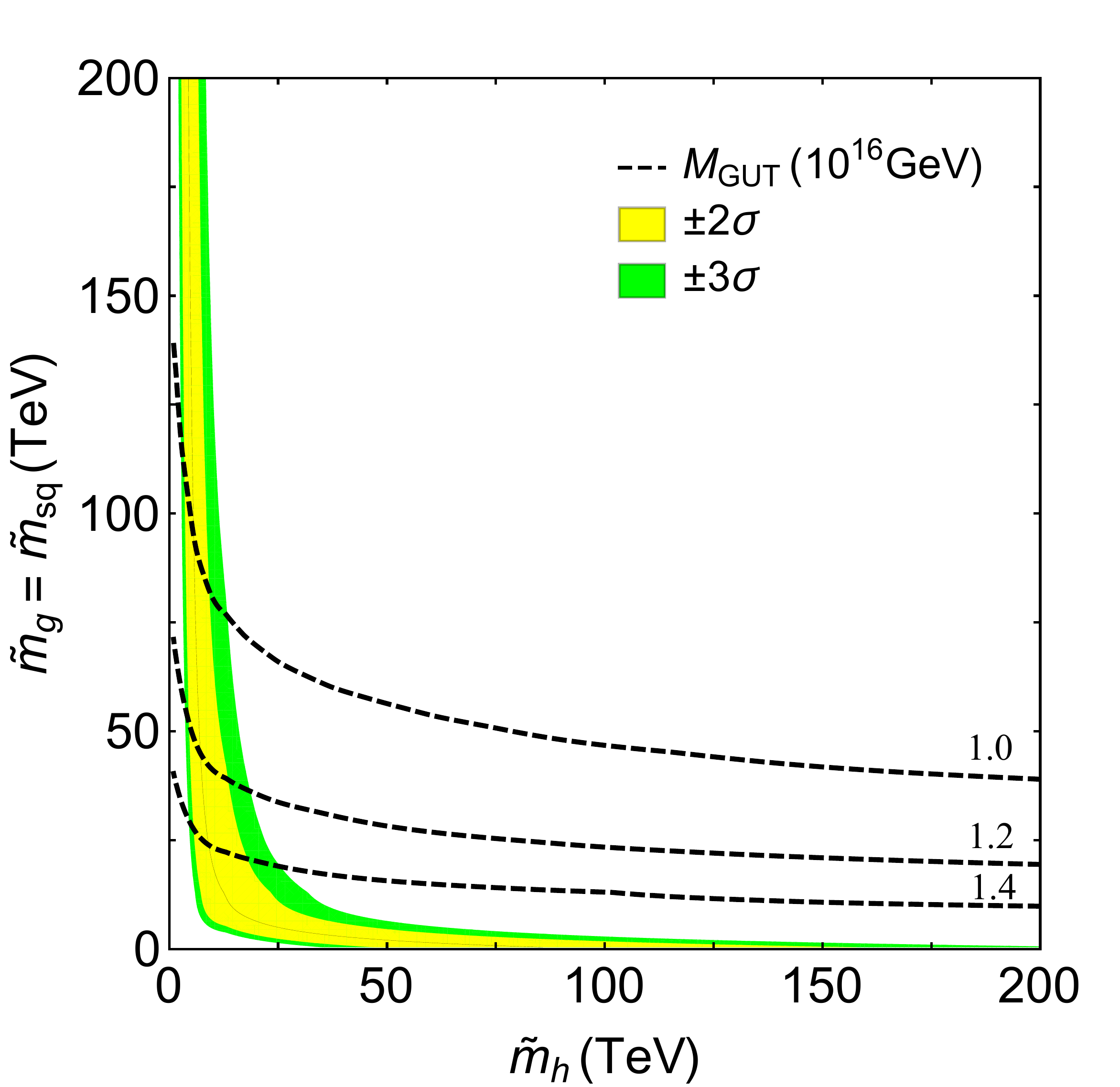}
\caption{The sections $\tilde{m}_{\rm g}=\tilde{m}_{\rm sq}$ of allowed SUSY mass spectrum are reported for two values of GUT threshold $\chi_\Sigma$ (yellow and green regions are 95\% and 99\% CL, respectively). The dashed black lines represent values of $M_{\rm GUT}$ in units of $10^{16}$~GeV, while the red line bounds from above the compatibility region due to proton decay limit.}
\label{fig:plots}
\end{figure}
Our study is performed by solving all the SM and MSSM Renormalization Group Equations (RGE) (see Ref.~\cite{Mihaila:2012fm,Mihaila:2012pz,Luo:2002ey,Chetyrkin:2012rz,Bednyakov:2012en} and \cite{Martin:1993zk} for SM and MSSM, respectively) up to 2-loop order through a numerical iterative method using a {\it Mathematica} code. Note that only the third generation Yukawa couplings provide a substantial contribution in the RGE, whereas the first and second generation ones can be neglected. Moreover, according to the decoupling theorem, we take into account SUSY and GUT 1-loop threshold relations at the mass of each new particle. In particular, we study the general case of several SUSY thresholds, {\it i.e.} the so-called {\it multi-scale approach}, in which the SUSY particles can have different masses.

The SUSY thresholds can be parametrized in terms of three mass scales only (see Ref.~\cite{Berezhiani:2015vea} for more details): the mass of higgsinos $\tilde{m}_{\rm h}$ (supersymmetric $\mu$-term), the mass of gluinos $\tilde{m}_{\rm g}$ ($F$-terms) and the mass of squarks $\tilde{m}_{\rm sq}$ ($D$-terms). Since the {\it SU(5) bottleneck} requirement implies that at $M_{\rm GUT}$ all gauginos must have the same mass and the same situation occurs for the particles that are allocated in the same GUT multiplet. Therefore, the masses of neutralinos and sleptons can be simply determined by the running.

Regarding the GUT thresholds, after the field $\Sigma$ ({\bf 24} adjoin representation of SU(5)) breaks SU(5) through a v.e.v., the mass $\tilde{M}_\Sigma$ of its fragments ($\left( {\bf 8}, \, {\bf 1} \right) \oplus \left( {\bf 1}, \, {\bf 3} \right)$ under $SU(3) \times SU(2)$) can be smaller than $M_{\rm GUT}$. Therefore, the GUT thresholds are defined by the parameter
\begin{equation}
 \chi_\Sigma \equiv \frac{M_{\rm GUT}}{\tilde{M}_\Sigma} = \frac{\sqrt {2 \pi \alpha_{\rm GUT}}}{\lambda_\Sigma}\, ,
\label{eq:chi_GUT}
\end{equation}
where $\alpha_{\rm GUT}$ is the gauge coupling at $M_{\rm GUT}$ and $\lambda_\Sigma$ is the trilinear self-interaction coupling of $\Sigma$. The condition $\lambda_\Sigma={\cal{O}}(1)$ ({\it naturalness} requirement) would imply $\chi_\Sigma \leq 10$, as it will be clear in the following. It is worth observing that also the color triplets of the two Higgses $H$ and $\overline{H}$, belonging to the fundamentals representations {\bf 5} and ${\bf \overline{5}}$, can have mass $\tilde{M}_T$ smaller than $M_{\rm GUT}$. However, in order to avoid a too fast proton decay via dimension-5 operator we assume $\tilde{M}_T \geq M_{\rm GUT}$.

For the sake of brevity, in Fig.~\ref{fig:plots} we report only the sections $\tilde{m}_{\rm g}=\tilde{m}_{\rm sg}$ of the 3D allowed regions of SUSY mass spectrum for two different values of $\chi_\Sigma$. One can see that the mass of higgsinos has an upper bound, which is an increasing function of GUT threshold. Moreover, the proton decay constraint (red line), which considerably affects only the region of small $\chi_\Sigma$, provides an upper bound for the mass of gluinos. It is worth observing that there exists an anti-correlation between $\tilde{m}_{\rm h}$ and $\tilde{m}_{\rm g}$ that allows us to look for the maximum $M_{\rm UB}(\chi_\Sigma)$ of the minima of compatible models $\{ \tilde{m}_{\rm h}, \, \tilde{m}_{\rm g}, \, \tilde{m}_{\rm sq} \}$.

\begin{figure}[t!]
\centering
\includegraphics[scale=0.245]{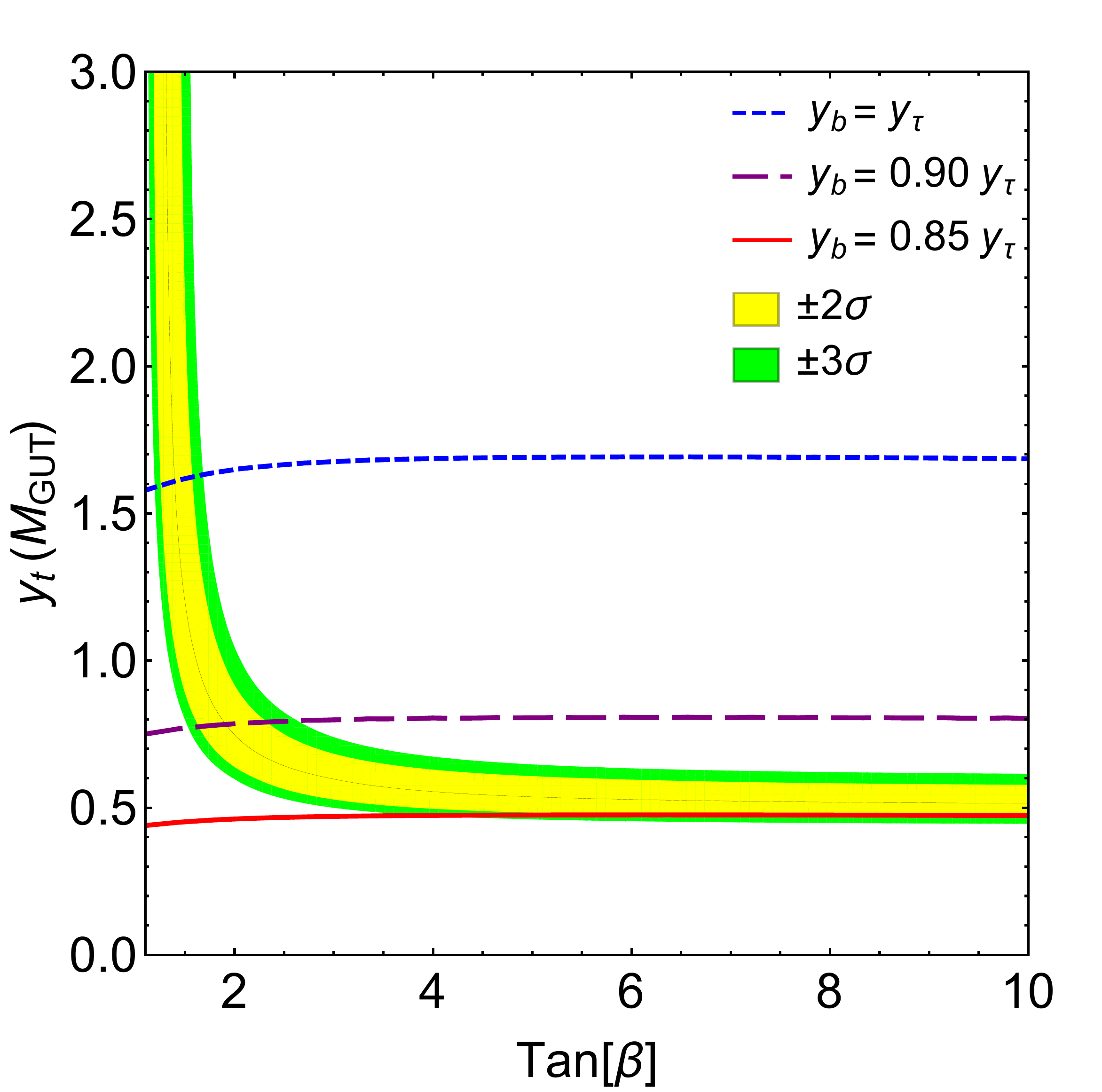}
\hspace{5.mm}
\includegraphics[scale=0.26]{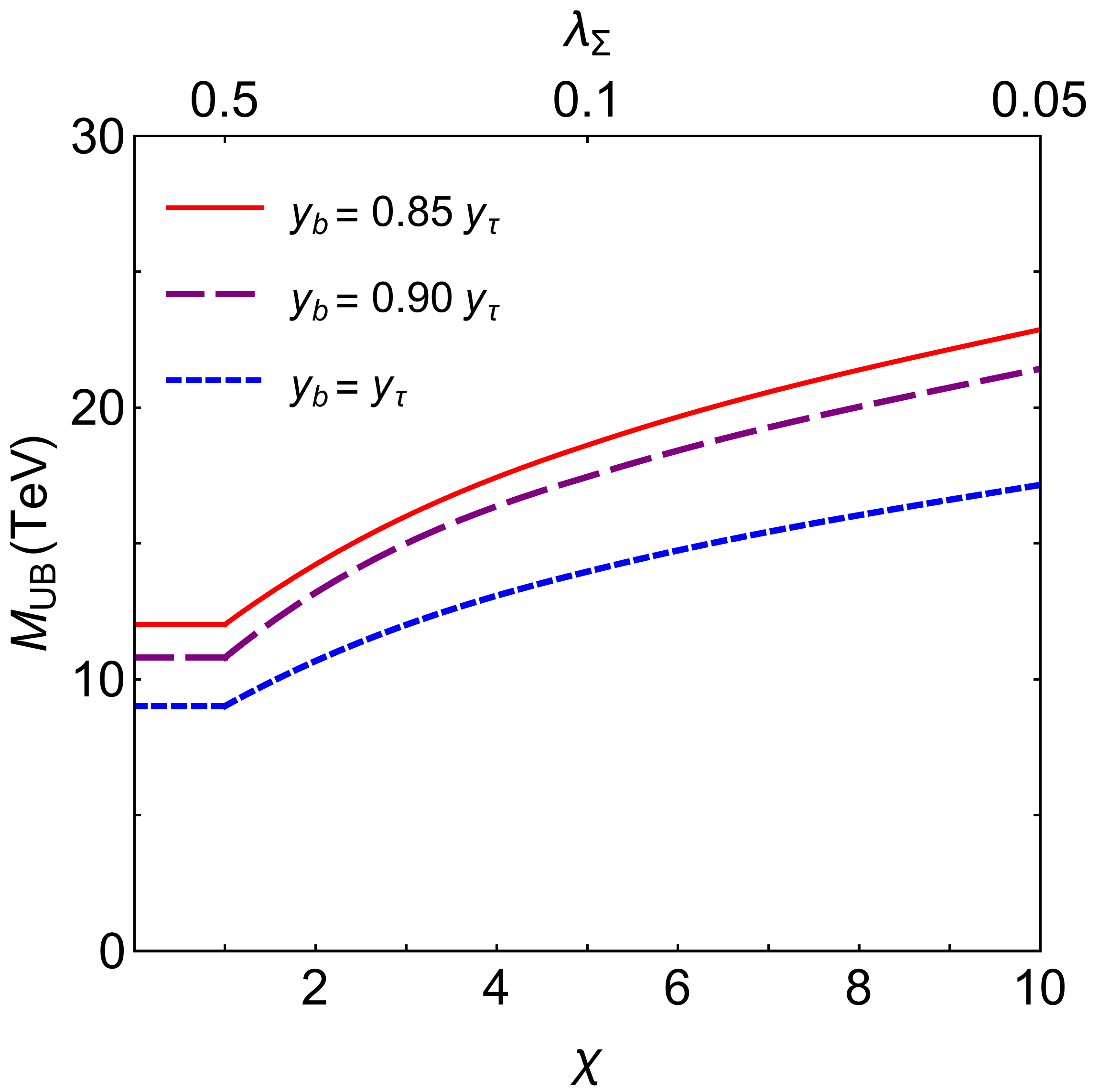}
\\(a) \hskip63.mm (b)\\
\caption{\label{fig:UB} (a) Compatibility region in the plane $\tan\beta$-$y_{\rm t}(M_{\rm GUT})$  (yellow and green regions are 95\% and 99\% CL, respectively). The lines bound from below the allowed region once that exact or partial $b$-$\tau$ unification at $M_{\rm GUT}$ is respectively assumed, according to Eq.~(\ref{eq:btau}). (b) The quantity $M_{\rm UB}$ as a function of $\chi_\Sigma$ and $b$-$\tau$ unification. On the upper part of the panel, the corresponding values of $\lambda_\Sigma$ (see Eq.~(\ref{eq:chi_GUT})) are reported.}
\end{figure}

The left panel of Fig.~\ref{fig:UB} shows the allowed region in the plane $\tan\beta$-$y_{\rm t}(M_{\rm GUT})$, where $\tan \beta$ is the ratio of the v.e.v.s taken by two MSSM Higgs doublets. The allowed region is differently bound from below by the requirements of exact or partial $b$-$\tau$ unification at $M_{\rm GUT}$, according to the relation
\begin{equation}
y_{\rm b}(M_{\rm GUT}) =y_{\rm \tau}(M_{\rm GUT}) \left(1 + {\cal O}\left( \frac{y_{\rm \mu}(M_{\rm GUT})}{y_{\rm \tau}(M_{\rm GUT})}\right)\right)\,.
\label{eq:btau}
\end{equation}
Finally, in the right panel of Fig.~\ref{fig:UB} we provide the quantity $M_{\rm UB}$ as function of GUT threshold $\chi_\Sigma$ and $b$-$\tau$ unification. As shown by the upper part of the plot, the {\it naturalness} requirement implies $\chi_\Sigma \leq 10$ since the value $\chi_\Sigma=10$ corresponds to $\lambda_\Sigma =0.05$. Hence, the {\it energy upper bound} for the observation of SUSY phenomenology results to be about 20~TeV. Such an energy corresponds to a bound for the mass of lightest gluino or higgsino. It is worth observing that a more accurate measurement of strong coupling at the electroweak scale, as well as a more stringent proton decay limit, would decrease this {\it upper bound}.

\section{Conclusions}

The SUSY-GUT paradigm is the most promising and elegant framework for the physics beyond the SM. After the LHC run I that has greatly constrained the SUSY phenomenology through a variety of direct and indirect searches, we have reanalyzed the room still remaining for SUSY-GUT. Considering SUSY and GUT thresholds, under some conditions ({\it SU(5) bottleneck} and {\it naturalness}) we have studied the bounds on SUSY mass spectrum coming from the requirement of agreement with the electroweak measurements, the masses of third generation fermions and the proton decay limit. We state that, if the SUSY-GUT paradigm holds, the mass of lightest gluino or higgsino cannot be larger than about 20~TeV. Such a limit is strongly dependent on the measurement of the strong coupling at the electroweak energy scale and on the proton decay limit.

\section*{\bf Acknowledgments} 

We acknowledge support of the MIUR grant for the Research Projects of National Interest  PRIN 2012 No. 2012CPPYP7 ``Astroparticle Physics".


\begin{thebibliography}{0}

\bibitem{Berezhiani:2015vea}
  \BY{Z.~Berezhiani, M.~Chianese, G.~Miele \atque S.~Morisi}
  \IN{JHEP}{1508}{2015}{083}
  [arXiv:1505.04950 [hep-ph]].

\bibitem{Nishino:2012ipa}
  \BY {H.~Nishino {\it et al.} [Super-Kamiokande Collaboration]}
  \IN{Phys. Rev. D}{85}{2012}{112001}
  [arXiv:1203.4030 [hep-ex]].

\bibitem{Hisano:2012wq}
  \BY{J.~Hisano, D.~Kobayashi \atque N.~Nagata}
  \IN{Phys. Lett. B}{716}{2012}{406}
  [arXiv:1204.6274 [hep-ph]].

\bibitem{Mihaila:2012fm}
  \BY{L.~N.~Mihaila, J.~Salomon \atque M.~Steinhauser}
  \IN{Phys. Rev. Lett.}{108}{2012}{151602}
  [arXiv:1201.5868 [hep-ph]].

\bibitem{Mihaila:2012pz}
  \BY{L.~N.~Mihaila, J.~Salomon \atque M.~Steinhauser}
  \IN{Phys. Rev. D}{86}{2012}{096008}
  [arXiv:1208.3357 [hep-ph]].

\bibitem{Luo:2002ey}
  {M.~x.~Luo \atque Y.~Xiao}
  \IN{Phys. Rev. Lett.}{90}{2003}{011601}
  [hep-ph/0207271].

\bibitem{Chetyrkin:2012rz}
  \BY{K.~G.~Chetyrkin \atque M.~F.~Zoller}
  \IN{JHEP}{1206}{2012}{033}
  [arXiv:1205.2892 [hep-ph]].

\bibitem{Bednyakov:2012en}
  \BY{A.~V.~Bednyakov, A.~F.~Pikelner \atque V.~N.~Velizhanin}
  \IN{Phys. Lett. B}{722}{2013}{336}
  [arXiv:1212.6829].

\bibitem{Martin:1993zk}
  \BY{S.~P.~Martin \atque M.~T.~Vaughn}
  \IN{Phys. Rev. D}{50}{1994}{2282};
  \SAME{78}{2008}{039903}
  [hep-ph/9311340].

\end{thebibliography}
\end{document}